# ICCTAFA 2019

## International Conference on Computer Technologies and Applications in Food and Agriculture

### July 11-12, 2019, Konya, Turkey

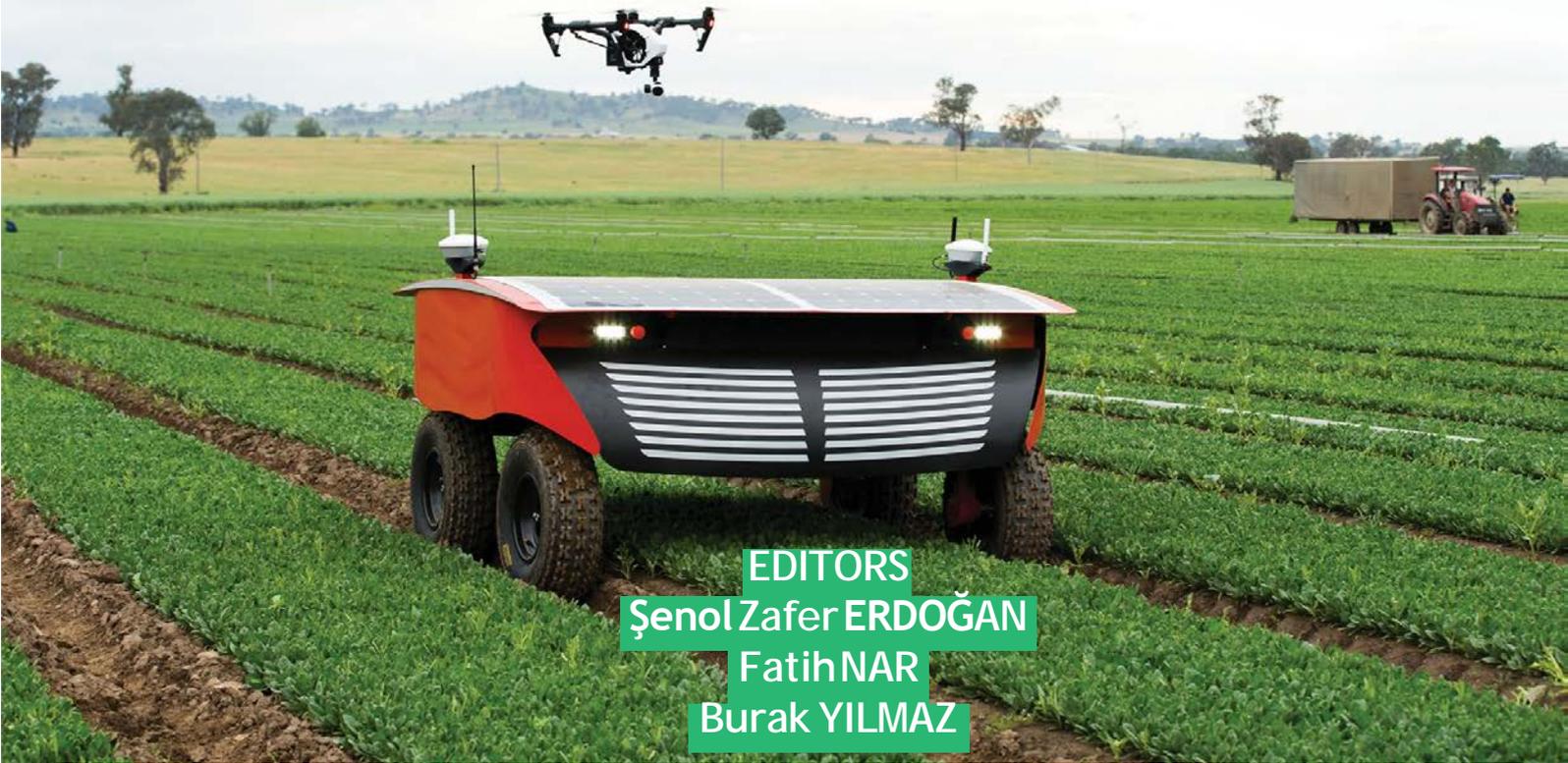

**EDITORS**
**Şenol** Zafer **ERDOĞAN**
Fatih **NAR**
Burak **YILMAZ**

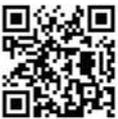
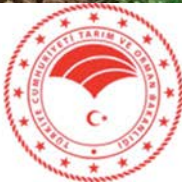
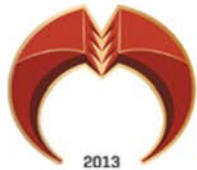
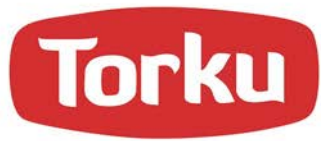
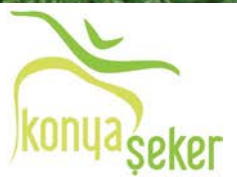

# An artificial intelligence and Internet of things based automated irrigation system

Ömer AYDIN
Dokuz Eylül University, Graduate School of Natural and Applied Sciences, Computer Engineering Department
omer.aydin@deu.edu.tr

Cem Ali KANDEMIR
cemalikandemir@gmail.com

Umut KIRAÇ
umut.kirac@outlook.com

Feriştah DALKILIÇ
Dokuz Eylül University, Faculty of Engineering, Computer Engineering Department
feristah@cs.deu.edu.tr

**ABSTRACT**: It is not hard to see that the need for clean water is growing by considering the decrease of the water sources day by day in the world. Potable fresh water is also used for irrigation, so it should be planned to decrease fresh water wastage. With the development of the technology and the availability of cheaper and more effective solutions, the efficiency of the irrigation increased and the water loss can be reduced. In particular, Internet of things (IoT) devices have begun to be used in all areas. We can easily and precisely collect temperature, humidity and mineral values from the irrigation field with the IoT devices and sensors. Most of the operations and decisions about irrigation are carried out by people. For people, it is hard to have all the real time data such as temperature, moisture and mineral levels in the decision-making process and make decisions by considering them. People usually make decisions with their experience. In this study, a wide range of information from irrigation field was obtained by using IoT devices and sensors. Data collected from IoT devices and sensors sent via communication channels and stored on MongoDB. With the help of Weka software, the data was normalized and the normalized data was used as a learning set. As a result of the examinations, decision tree (J48) algorithm with the highest accuracy was chosen and artificial intelligence model was created. Decisions are used to manage operations such as starting, maintaining and stopping the irrigation. The accuracy of the decisions was evaluated and the irrigation system was tested with the results. There are options to manage, view the system remotely and manually and also see the system's decisions with the created mobile application.

**ÖZET**: Dünyadaki temiz su kaynaklarının günden güne azalması göz önüne alındığında temiz su ihtiyacının arttığını görmek zor değildir. Temiz içme suyu aynı zamanda sulama için de kullanılır bu nedenle temiz su israfı azaltma süreci planlanmalıdır. Teknolojinin gelişmesi, daha ucuz ve daha etkin çözümlerin ortaya çıkması ile birlikte, sulama verimliliği artmakta ve su kaybı azalmaktadır. Özellikle, Nesnelerin İnterneti cihazları (IoT) tüm alanlarda kullanılmaya başlanmıştır. IoT cihazlar ve sensörler ile sulama alanından sıcaklık, nem ve mineral değerlerini kolayca ve hassas bir şekilde toplayabiliriz. Günümüzde sulama ile ilgili işlem ve kararların çoğu insanlar tarafından yürütülmektedir. Karar verme sürecinde sıcaklık, nem ve mineral seviyeleri gibi birçok gerçek zamanlı veriye sahip olmak ve bunları dikkate alarak karar vermek insanlar için zordur. İnsanlar genellikle kendi deneyimleriyle karar alırlar. Bu çalışmada, IoT cihazları ve sensörler kullanılarak sulama alanından geniş bir veri toplanmıştır. IoT cihazlarından ve sensörlerden toplanan veriler, iletişim kanallarından sunucuya aktarılır ve MongoDB üzerinde saklanır. Weka yazılımı yardımı ile normalizasyon işlemleri yapılan veriler öğrenme seti olarak kullanılır. Denemeler sonucunca yüksek başarı oranına sahip karar ağacı (J48) algoritması seçilmiş ve yapay zeka modeli oluşturulmuştur. Kararlar, sulamayı başlatmak, sürdürmek ve durdurmak gibi işlemleri yönetmek için kullanılmıştır. Kararların doğruluğu değerlendirilmiş ve sulama sistemi sonuçlarla test edilmiştir. Oluşturulan mobil uygulama ile sistemi uzaktan ve manuel olarak yönetmek, görüntülemek ve ayrıca sistemin vermiş olduğu kararları görebilmek için seçenekler vardır.

**Anahtar Kelimeler:** Yapay zeka, Nesnelerin İnterneti, Sulama, Makine öğrenmesi, Sensörler

# INTRODUCTION

Water is one of the most basic items that society needs. It plays a vital role in the development of the nutrients that the living beings need to survive, in the functioning of the body's functions and in the life cycle of the world. Using the water more efficiently and at the most appropriate level is critical to the future of the earth.

Factors such as global warming lead to the reduction of clean water resources in the world and this situation is getting serious day by day. The scarcity that will arise in the event of losing clean water resources will put the vital continuity of the creatures at risk. Years of efforts to prevent global warming continue, but the desired result has not been achieved.

Wasting water while the water resources on the earth are decreasing makes the situation more difficult. Inadequate or excessive irrigation in the irrigation sector constitutes a significant amount of wasted water. On the other hand, inadequate areas of inefficient vegetation cannot adequately meet the oxygen needs of living things. In addition, agricultural products are damaged in this situation and reduce harvesting efficiency. This leads to both water and nutrient loss.

Today, irrigation is mostly proceeded with human decision. This situation causes people to be inadequate at some decision points. For example, the time required for product to be irrigated is determined by people's estimation and experience. However, when these determinations are made, the number of data taken into consideration is limited and it is not certain. It is unrealistic to expect a person to consider whether the most appropriate value for the product, such as weather, soil moisture, soil temperature, minerals in the earth, has been reached when deciding on the irrigation. For this reason, very accurate decisions cannot be made.

Irrigation area owners and producers are unable to follow the real-time situation in their area and cannot change these situations with remote access. This poses a problem when the producers can not physically reach it. As a solution to this, extra human power is being used; therefore, the extra cost is incurred.

**Motivation**

The Internet of things (IoT) sector is evolving with computers being more involved in human life. With IoT applications, it is possible to include computers in the field of irrigation. Taking into consideration the numerical data indicating the current situation in the irrigation decision system, decisions will be much more accurate than decisions to be made according to the individual's own personal judgments. For this reason, computerization in irrigation systems will increase water, oxygen and production efficiency in this area by taking decisions to more accurate results.

The studies that are being investigated generally provide the users with automatic retrieval of data via e-mails and present them to the users, reminding them of their watering and helping them to make decisions. These systems do not use machine learning, artificial intelligence methods so they are not the intelligent irrigation systems.

**Contribution**

Thanks to the intelligent irrigation system, the products will always be irrigated on time, so that the products will get the best results. Since the amount of irrigation is fully programmable, the product will not be given more than necessary so that a large amount of water will be saved. The cost and the responsibility for the producers will be reduced because the computer control and devices are used in the irrigation processes rather than human power. Vital values of the application area can be intercepted, which can be monitored from mobile devices, which are always available to those who own the area. Remote access is also possible.

The rest of this paper is organized as follows. Section 2 includes previous and related works about this paper's subjects. In Section 3, proposed work is explained. In Sections 4 and 5, materials and methods used for proposed system are explained in detail. Section 6 gives experimental results of performance, test, evaluation and discussions about them. Finally, we conclude the paper in Section 7.

# RELATED WORKS

As the developing technology in many areas, there are many applications in the field of agriculture. Significant studies have been carried out in recent years, especially on the monitoring of soil status and automatic implementation of irrigation activities. In these studies, sensors and micro controllers have been used in general.

In a study by Dhiman et al. [1], a decision management system on drip irrigation that considers when the water will open and close has been presented. This study shows the practical prevention of water waste in greenhouse/closed area growing by improving irrigation water usage 90%. In another study, it was aimed to perfect irrigation in a certain region by automating the irrigation system [2]. Information about temperature, humidity and soil status is transmitted to a microcontroller. The microcontroller transmits commands to a motor to run or stop. The information in the microcontroller is sent to a liquid crystal display (LCD) screen via global system for mobile communications (GSM) as short message service (SMS). In addition, farmers are informed with SMS about the exact conditions on the field.

In a study, that aims minimizing water consumption in agricultural irrigation processes, soil moisture and temperature sensors are located in the root zone of plants [3]. Distributed wireless network is used and the data collected from the taps are shared to a web application with the help of the triggering actuator. An algorithm was developed with the aim of controlling the amount of water, with programmed micro-base passages and threshold values of temperature and soil nudity. The system was tested in a 136-day sage field and water savings of up to 90% were observed. Pavithra and Srinath used general packet radio service (GPRS) feature of mobile phone as a solution for irrigation control system and delivered the current field conditions to the user in the form of SMS via GSM [4]. They proposed an android phone-controlled irrigation system to maintain uniform environmental conditions at all the places in large farmhouses, which make very difficult to maintain the uniformity manually. Ergün et al. used Arduino Uno, a soil moisture sensor, a solenoid water valve and two XBee devices and activated the water valve via terminal screen [5]. The ZigBee network was used for remote control of the water valve and dripping method was used as irrigation method. In another study, an automatic irrigation system using 32-bit ARM7TDMI microcontroller and GSM infrastructure was developed [6]. With the aid of the sensors, the situation in the vicinity is determined and the relevant information is displayed on the LCD to inform users.

Krishna and Aswini aimed to measure soil status in real time by improving irrigation system by the real-time operating system (RTOS) [7]. Using this system, the status of the area can be detected with the help of the sensor and relevant information is displayed on an LCD for the users. Zhang et al. [8], proposed an agricultural irrigation system by combining data transmission unit (DTU), wireless radio frequency (RF) unit and micro controller. In data transmission process, GRPS network is used. They powered the entire system by solar energy. In a recent study, an irrigation system has been designed by using a wireless sensor network that includes a wireless sensor unit and a wireless information unit [9]. This network governs the amount of water by following the temperature, ambient humidity, and soil nematode and soil pH. The system produces power through photovoltaic panels.

In another recent study, a system consists of Arduino micro controller, DS1302 module, water pump, potentiometer, and Arduino motor drive circuit, has been developed [10]. The system does not have network connectivity, sensor data and mobile control options.

The studies that are being investigated generally provide automatic retrieval of data via short messages or e-mails to remind the user for watering and helps them to make decisions. These systems do not use machine learning, artificial intelligence methods so they are not the intelligent automated irrigation systems.

## PROPOSED WORK

The project will be applicable to every area where the irrigating is made. When used in agriculture, it will ensure the development of the products in the most favourable conditions and thus increase the overall efficiency of the application area. In addition, financially, the producer will be a great asset. Contrary to the timely systems currently used in landscape and urban irrigation, water consumption will not be constant and will prevent water wastage taking into account rain and other conditions. It will also keep the view in the landscape areas at the desired level, avoiding rotting or fading. Worm farms will also provide the most appropriate solution for worm farming-critical water needs, which will be reflected positively in other sectors such as fisheries.

In Figure 1, the general architecture of the proposed work is presented. The data received via the sensors are collected by WeMos D1 Mini microcontroller [11] and sent to the message queuing telemetry transport (MQTT) server [12]. The MQTT Server also sends the obtained data to the mobile application for user monitoring; to the MongoDB database [13], for storing; and to the machine-learning algorithm for deciding the action. The irrigation decision resulting from the machine learning algorithm is transmitted to the WeMos D1 Mini. The device can activate the irrigation system, according to this decision.

The data received by various sensors at different points are transmitted to the server by the WeMos D1 Mini integrated card. This card is one of the Arduino derivatives that we use for the microcontroller role and it has Wi-Fi module on it. The sensors used consist of soil moisture sensor, air humidity and temperature sensor, and rain sensor. MQTT was chosen as the protocol used to send data received from these sensors. The reason for choosing this protocol is that, it requires much less processor power and bandwidth than other protocols to operate because of its IoT devices. In addition, through Publish / Subscribe architecture, it is possible to broadcast by a certain number of devices, while simultaneously listening to the desired devices from certain channels. The agent (broker) that provided this architecture was selected as Mosquitto Broker and installed on Linux server. Through this server, the developed mobile application is communicated and the data is displayed and control of the system is presented to the user in a flexible way. In addition, this microcontroller listening to the server; controls the submersible engine which will activate the irrigation system.

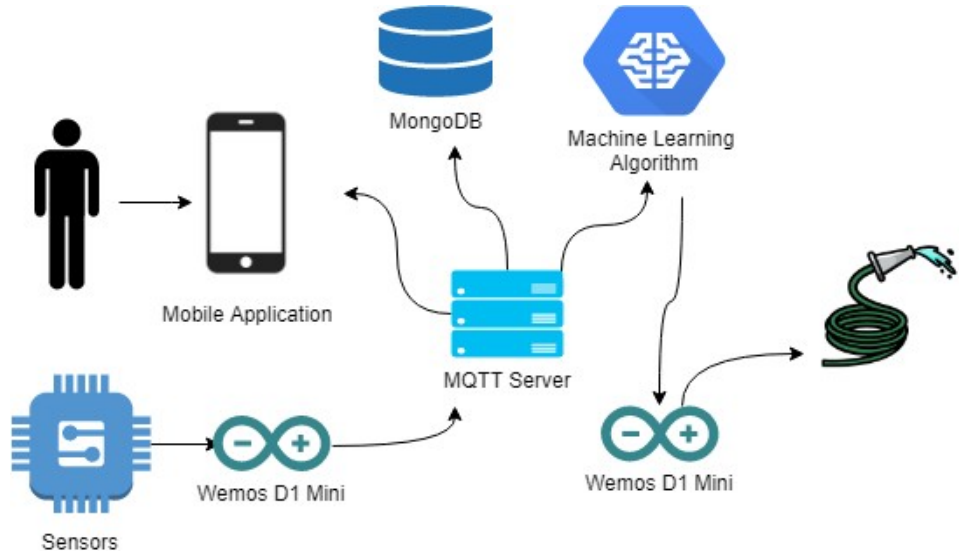

**Figure 1. General Look of Proposed Work**

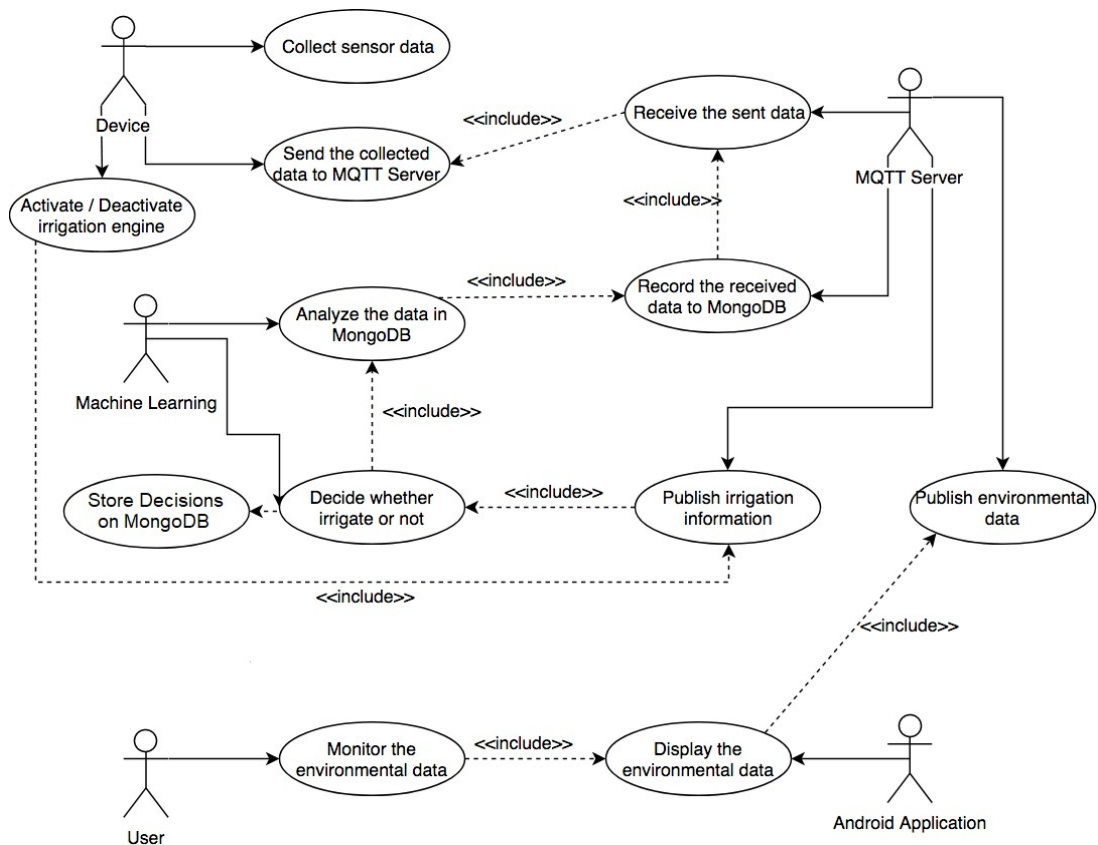

**Figure 2. Use Case Diagram**

In Figure 2, use case diagram has five actors. Device collects environmental data and sends them to the MQTT Server. MQTT Server records the data to the MongoDB and then machine-learning algorithm analyses the data with previously saved decisions and decides whether irrigate or not. This decision is saved to the database for further decision making process. The device controls the irrigate engine according to the decision of machine-learning algorithm. These decisions are also stored on database for the further evaluations. In addition, user monitors the system and decisions by the android application.

Intelligent irrigation system devices will be deployed at certain distances and locations, depending on the size of the irrigation environment and the crop grown. Energy requirements can be met by the power sources in the environment, and the devices to be connected will be distributed depending on this location. The circuit elements will be stored in a box that will be designed to be minimally affected by ambient conditions. The mobile application to be developed will have a user friendly and easy to understand interface by graphing important data.

**MATERIALS**

Hardwares such as soil moisture sensor, humidity and temperature sensor, water engine circuit, WeMos D1 mini and rain sensor are used in this study. Necessary data is picked and sent to server by these hardwares. Connections, structures and other details are given in the following subsections.

**FC-28 soil moisture sensor**

The FC-28 Soil Moisture Sensor (Figure 3) is one of the most successful and inexpensive sensors. General working logic; is to give information about how damp the environment is due to the difference in tension between the electrodes on the sensor due to the resistance of the environment such as soil or liquid. If analog data is being received from the sensor, the received data is converted to an integer value between 0-1023. The data on this scale may be numerically small and the environment may be so moist. As a result of the tests, the value seen in the tap water is around 120-180, while in a pot with moist soil it is 400-500; in the case of a pot with dry soil, it is in the range of 700-800. If the environment is air, the values are close to 1000 [14].

With a dynamic implementation algorithm, multiple FC-28 Sensors can be added to the system depending on the microcontroller ports. The data from these newly added sensors can be included in the algorithm. For this operation, the data may be averaged or subjected to normalization. In the prototype study, only one FC-28 Sensor was used. It is positioned in the root zone of the plant.

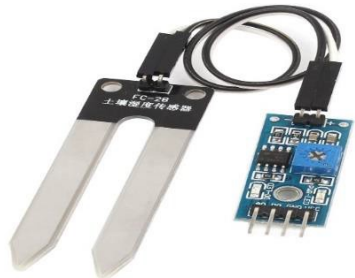

**Figure 3. FC-28 Soil Moisture Sensor**

**DHT11 humidity and temperature sensor**

DHT11 humidity and temperature sensor (Figure 4) is a cheap sensor that can do a lot of work easily and can be easily found on the market. Temperatures between 0 and 50 degrees Celsius can be measured with a margin of error of 2 degrees Celsius. It can also measure moisture content between 20% and 80% with a 5% error margin. The sensor, which can operate between 3 and 5 volts, is well suited for microcontrollers such as WeMos, which can generate voltage at these intervals [13].

It simply contains the humidity sensing component, negative temperature coefficient (NTC) temperature sensor (thermistor) and an integrated circuit (IC). The change in conductivity of the underlying layer which can hold moisture and the difference in resistance between the electrodes are measured. This measurement gives moisture. It is processed by IC and made ready for reading by the microcontroller. When measuring the temperature in a similar way, a decrease in resistance is observed as the temperature increases through the NTC, which is processed by the IC.

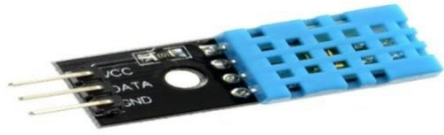

**Figure 4. DHT11 Humidity Sensor**

The sensor is used as a single component in the prototype and it is concluded that the ambient temperature is sufficient to measure the humidity of the air. In the implementation, the environment temperature and the percentage of ambient temperature that this sensor digitally sends are transmitted to the server to participate in the account.

**Rain sensor**

The rain sensor that is used (Figure 5) is a simple and inexpensive sensor that reports the voltage difference caused by the contact of the parallel conductive lines on the rain water with rain on the environment. The sensor is considered a suitable component for the prototype because it is exposed to continuous rain and the oxidation time is long and it can operate at 5 volts [16].

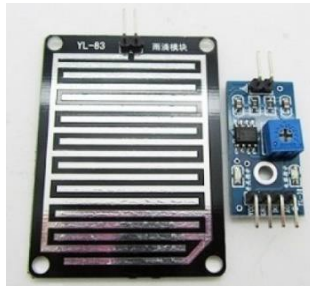

**Figure 5. Rain Sensor**

The only rain sensor in the prototype is used to cut off the watering when it rains. Instead of commenting on the Internet by air, informing the existence of rain specifically about the plant has made it an indispensable component for the prototype.

**WeMos D1 mini**

A microcontroller was needed at the stages of collecting, editing and sending data from various sensors to the server in the proper format. The used devices had to be able to supply required voltages to feed their digital pins and Wi-Fi module to connect internet. We decided to use the WeMos D1 Mini Pro microcontroller after comparing the needs, size and cheapness.

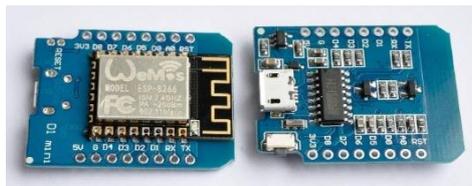

**Figure 6. WeMos D1 Mini**

The WeMos D1 Mini Pro (Figure 6) meets the requirements internally due to its built-in ESP8266 Wi-Fi module, 11 digital I/O pines, 80 MHz/160 MHz clock speed and 34.2x25.6 mm size.

The microcontroller, which is easily programmable with Arduino IDE, publishes the data to the topic specified in the MQTT server, collecting data from the sensors every 5 minutes. At the same time, it can be programmed to read data from specific topics and to act according to the situation.

**Water engine circuit**

If the data are analysed and it is decided to water according to the algorithm to be used, the prototype pot must be watered in various ways. For this, a motor was needed to draw water from many water sources. For the prototype, we decided to use the submersible engine (Figure 7) that is immersed in a water source.

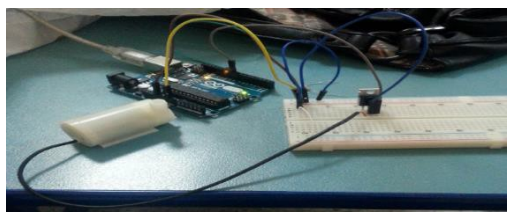

**Figure 7. Water Engine Circuit**

The used submersible motor requires 5 Volt tension, sucking water from the hole in the lower part and discharging water from the notch in the upper part. It is a pot for waterproofing the prototype. With the aid of a triac, the water engine can be started and stopped according to the signal sent digitally.

## METHODS

In this section, collection of data to be used to train the machine-learning algorithm from the environment, data set creation and normalization, selection and implementation of appropriate machine learning algorithm are explained.

**Collecting the data from environment**

After completing the necessary hardware installations in our smart irrigation project, we switched to the software part. Through the used sensors, data such as temperature, humidity and soil moisture and rain condition were collected from the environment and these data were sent to the server for storing in the database. This process was implemented with software developed by WeMos.

Firstly, the necessary libraries were determined and implemented. The ESP8266WiFi library is a Wi-Fi module and is required for the ESP8266 integrated on WeMos. Using this library, Wi-Fi connection is realized via ESP8266. The PubSubClient library is used to exchange data with the MQTT server. The dht11 library is required to collect temperature and humidity data from the environment.

Once the required libraries have been defined, the necessary constants are defined. These constants consist of the Wi-Fi information to be connected and the IP address of the server to exchange data with. Then a connection request is sent to the Wi-Fi with the next code and a loop is created until the connection is made. Also a loop was created until the first connection to the server was established. Then a clientID was created. After the clientID is created, the connection request to the server is sent with the next command. This command sent the clientID, username and password information to the server.

After Wi-Fi and server connections are done, functions of reading data from sensors and writing data to server are coded. First, the data acquisition codes of the sensors according to their analog and digital readings are written. The soil moisture value, rain condition, air temperature and humidity data are obtained from the pins to which the sensors are connected, depending on whether they are analog or digital. The read data was first converted to a string before being written to the server. Briefly, this code written on the hardware side collects data from the sensors. Wi-Fi and server connections were made and the collected data was written to the server.

**Dataset creation and normalization**

In our intelligent irrigation project, we developed an artificial intelligence for the decision of irrigation on the data collected from the environment through the sensors of the data required for irrigation decision.

The first step in the development of artificial intelligence was the creation of a training set and the realization of machine learning. The first thing that is important for the accuracy of artificial intelligence decisions to be high is that the training set to be taught to the machine is strong. The variety of data on the trainings and the many different conditions will strengthen artificial intelligence. This information was paid attention when the training set was created in the project.

In order to create the training set, the sensors were installed on the ground and data collected from the environment for about 3-4 months. It was noted that normalization procedures were performed on the collected data and a strong training set was created. Necessary actions were taken on the data we collected in the scope of the studies for normalization operations. First, the collected data was extracted. At this time, the data was selected and transferred to a new dataset. Redundancy was reduced by selecting the data for certain periods.

Then the data in the generated dataset was cleaned. Noisy data was normalized. Null data was deleted in some cases, but in some cases it was corrected using the values of the nearest neighbours. (K-NN algorithm)

After the created data set has been cleaned, a desired data set has been obtained in terms of data diversity and correctness. After this step, when artificial intelligence decided, dominance adjustment was made between soil moisture, air humidity and temperature, and rain condition data. In machine learning, the highest value of the data in each data row is the dominant value. For this reason, when artificial intelligence decides that soil nematodes value in the range of 400-800, rain condition 0-1, soil nematode considers much more than rain condition data. This affects the accuracy of artificial intelligence decisions negatively. For this reason, the normalization process of each value has been done.

Several methods were used at this stage. These methods are as follows; Z-Score, Max-Min, maximum number of divisions. It has been observed that the Z-Score method, which is said to have the highest accuracy in theory after applying each of these methods separately, gives the most accurate results in practice. Thus, it was decided to implement the Z-Score method. Finally, the generated dataset was normalized with Z-Score and the dataset was finalized and prepared for use in machine learning as training set.

**Machine learning algorithm selection and implementation**

In the process of selecting the machine-learning algorithm, Weka, a popular and successful tool, was utilized. Put simply, it is a tool that is easy to compare, testing readily created implementations of known algorithms on the market with custom data sets like in Figure 8 that can be prepared by the user. Normalization on the prepared dataset presents many numerical and graphical data such as the rate of trisection of these errors by selecting many classified algorithms and specifying training and testing clusters.

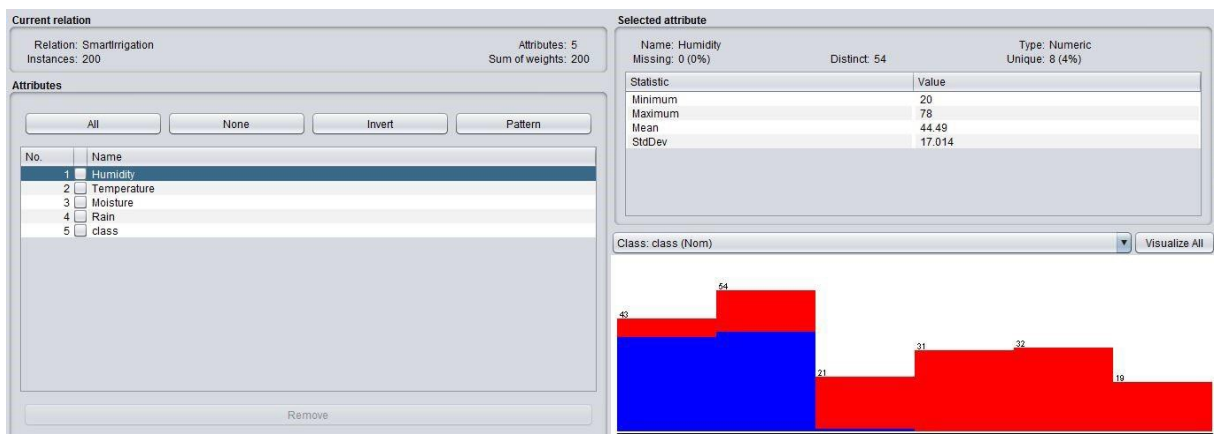

**Figure 1. Training Set, Before Normalization**

Through the training set we created in our project, the values were set between 0 and 1 with the help of the normalization function in Weka, so that we tried to eliminate the errors that would arise from the numerical difference between the features as seen on Figure 9. Afterwards, many algorithms were chosen from the normalized data set, and the most suitable machine-learning algorithm was tried to be selected by taking into account the error rates obtained from these algorithms, the ability to create models, the ease of implementation, and the resource adequacy of the algorithm.

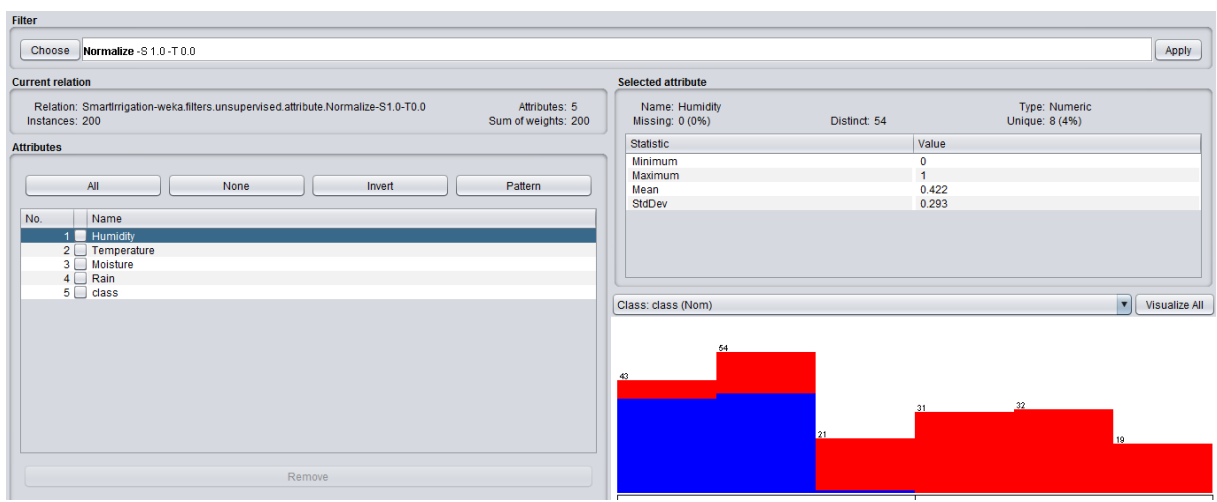

**Figure 9. Training Set, After Normalization**

```
Time taken to build model: 0 seconds

=== Stratified cross-validation ===
=== Summary ===

Correctly Classified Instances         196                98      %
Kappa statistic                          0.9571
Mean absolute error                      0.0338
Root mean squared error                  0.1407
Relative absolute error                  7.1948 %
Root relative squared error             29.062  %
Total Number of Instances              200

=== Detailed Accuracy By Class ===

                 TP Rate  FP Rate  Precision  Recall  F-Measure  MCC    ROC Area  PRC Area  Class
                 0,960    0,008    0,986      0,960   0,973      0,957  0,965     0,963     1
                 0,992    0,040    0,976      0,992   0,984      0,957  0,965     0,960     0
Weighted Avg.    0,980    0,028    0,980      0,980   0,980      0,957  0,965     0,961

=== Confusion Matrix ===

   a   b   <-- classified as
  72   3 |   a = 1
   1 124 |   b = 0
```

**Figure 10.** J48 Algorithm Results on Real Data

In Weka algorithms normalized data were tested in 10-fold cross-validation test mode. Algorithms with the highest accuracy are examined. Because of its high success rate (98%) and F-Measure value, both modelable tree structure and fast operation, J48 algorithm (Decision Tree) was suitable. The results can be seen on Figure 10.

The machine-learning results generated using the training set can be saved as a static model. In addition, in C #, all algorithms and models written in Weka can be used with the help of the Weka library. In our project, a static model was created by the J48 algorithm, which is fed with the training set prepared by using this library and the lid. It was decided to start, continue or stop irrigation by taking instant data as a parameter.

## 6. Experimental Results

In this section, the creation of a test environment for testing the contribution of the machine-learning algorithm to the project and the analysis of the results of the tests made are explained.

### 6.1. Setting up the test environment

Since our work is very weak in real-world time flow testability, seeing the value of the irrigation command leads to a considerable waste of time. Instead, a simulation was created to simulate the entire environment. In this simulation, the user enters the desired environment data. This environmental data is divided into four sections as air humidity, temperature, soil moisture and rainfall. The user transfers this data to the Topic created for testing in the MQTT Server used by any tool and is collected by an application that continuously listens to this Topic and runs a Callback method when new data arrives. Server; It has the ability of listen to the desired MQTT Server, broadcast to the desired Topic on this server, and make decisions using the "J48 - Decision Tree" algorithm, one of the Machine Learning algorithms that it has.

For testing; MQTT Lens (see Figure 11) software was used as a tool to publish / subscribe to MQTT Server. This simple to use tool connects to the desired MQTT server from the desired port. It is then possible to subscribe to the desired Topic via the server it connects to, or to send the desired message to a specific Topic.

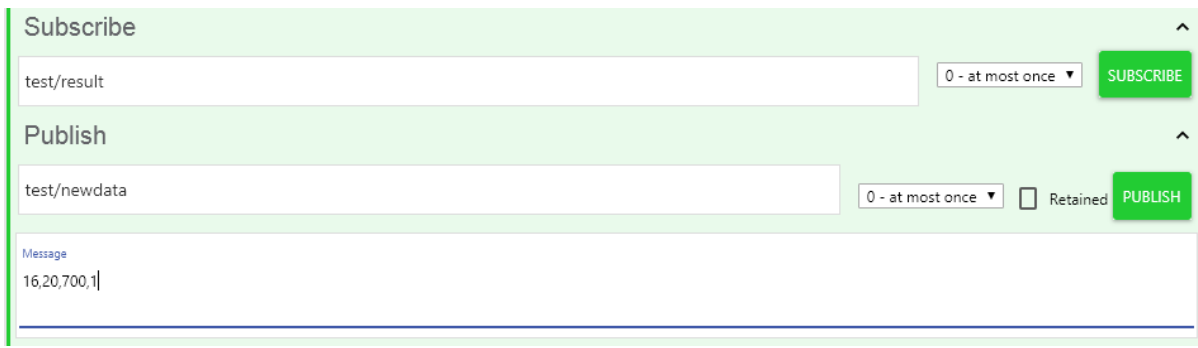

**Figure 11.** The tool used for testing, MQTTLens.

The implemented application is used to see the test results. When this application is run for the first time, it creates a Decision Tree Model using the training set which is provided manually. In addition, this application listens to the "test / newdata" Topic created for the test, parse the incoming data when a new data arrives, and returns the values as parameters for making a decision in the Decision Tree Model. Depending on the structure of the data set used in our work, the return data can be "0 - Do not irrigate" or "1 - Start irrigating". The resulting data is again shared in the "test / result" Topic that is created for the test.

**Table 1**. Data tests and results.

| DATA | PREDICTED | ACTUAL | RESULT |
|---|---|---|---|
| 78,9,485,1 | 0 | 0 | TRUE |
| 58,10,447,1 | 0 | 0 | TRUE |
| 35,11,748,0 | 1 | 1 | TRUE |
| 35,18,775,0,1 | 1 | 1 | TRUE |
| 24,29,898,0,1 | 1 | 1 | TRUE |
| 35,22,852,0,1 | 1 | 1 | TRUE |
| 38,35,837,0,1 | 1 | 1 | TRUE |
| 45,31,565,1,0 | 0 | 0 | TRUE |
| 64,38,564,1,0 | 0 | 0 | TRUE |
| 55,40,556,1,0 | 0 | 0 | TRUE |
| 66,32,411,1,0 | 0 | 0 | TRUE |
| 51,35,552,1,0 | 1 | 0 | FALSE |
| 65,20,488,1,0 | 0 | 0 | TRUE |
| 30,30,838,0,1 | 1 | 1 | TRUE |
| 35,11,748,0,1 | 1 | 1 | TRUE |
| 39,14,682,0,0 | 0 | 0 | TRUE |
| 31,13,628,0,0 | 0 | 0 | TRUE |
| 24,14,694,0,1 | 1 | 1 | TRUE |
| 31,19,796,0,1 | 1 | 1 | TRUE |
| 70,14,415,1,0 | 0 | 0 | TRUE |
| 69,20,536,1,0 | 0 | 0 | TRUE |
| 60,19,506,1,0 | 0 | 0 | TRUE |
| 60,14,522,1,0 | 1 | 0 | FALSE |
| 72,9,455,1 ,0 | 0 | 0 | TRUE |
| 55,10,427,1,0 | 0 | 0 | TRUE |
| 52,10,536,1,0 | 0 | 0 | TRUE |
| 71,12,482,1,0 | 0 | 0 | TRUE |
| 64,13,583,1,0 | 0 | 0 | TRUE |
| 53,9,561,1, 0 | 0 | 0 | TRUE |
| 50,9,440,1, 0 | 0 | 0 | TRUE |

Thus, the user can test the desired number of data using the model created at the beginning and see what the result might be. Thanks to this, every step of the operation of the system can be tested.

**Testing results for machine learning part**

Tests were conducted on the data interval experienced during the study using the generated test environment. As a result of 30 tests, 94% success was achieved as given in Table 1.

The "DATA" column is in the format [Humidity, Temperature, Soil Moisture, is Raining]. They created completely manually based on previous measurements. The "PREDICTED" column shows the results predicted by the Machine Learning algorithm. The "ACTUAL" column is an experience-based irrigation decision that is experienced in similar data. These decisions are subjective decisions, assessed according to values such as the colour of the leaves, the establishment of the soil, it is not certain that all is correct. The "RESULT" column shows whether the values estimated by the machine learning algorithm and the evaluations using human experience overlap.

## DISCUSSION AND CONCLUSION

Agriculture has been a topic of great importance in human life since its inception and has developed as much as every day. People in the field of agriculture have constantly tried various methods and aimed to get the best results using the most efficient methods. The most important points of attention in agriculture were water and harvest. This situation has made irrigation methods extremely important.

The common point of all past and present irrigation methods is human dependency. Irrigation decisions and quantities depend on people and they have important times during the day. Correct irrigation depends on the environmental conditions and the needs of the products to be irrigated. The various factors that come into play at this point make it difficult for people to make the right irrigation decisions and irrigation is not done properly. Many irregularities arise as a result of improper irrigation operations, most of which are water wastage, inefficient crops, lost agricultural land, cost and labour loss.

The way to make the right irrigation and increase the yield is through machines and artificial intelligence. Thanks to irrigation equipment and artificial intelligence, proper irrigation can be provided by taking all the necessary factors into consideration. In this study, environmental conditions including humidity, temperature, moisture and rain have been taken into account. For this reason, humidity, temperature, moisture and rainfall conditions were measured from the environment through the sensors used in the field to be irrigated. Measurements made with specific periods were sent to the MQTT Server via the WeMos D1 Mini and the MQTT Server stored the received data in MongoDB. At the end of 4 months, a training set was created on the data set in MongoDB and this training set was tested on various artificial intelligence algorithms via Weka. As a result of the examinations, decision tree (J48) algorithm with the highest accuracy was chosen and artificial intelligence model was created. The artificial intelligence algorithm is coded by using C# programming language. The algorithm makes an irritation decision and returns the decision to the MQTT Server. The decisions are also stored in the database. The MQTT Server sends a command to another WeMos D1 Mini in the irrigation environment, and the WeMos D1 Mini operates the irrigation motor according to command it receives. This process repeats the cycle with adjustable periods and the soil is irrigated when necessary. Thanks to usage of artificial intelligence in irrigation decision, we obtained a 94% success in irrigation process and prevented water wastage, decreased the cost and loss of work power. We believe that this system will also increase the product efficiency. Human can control and monitor system with the mobile application. The system can be used in fields or areas of application with varying dimensions. There is no limit for the size of area. This can only be achieved by increasing the number of data collection units in which the sensors are installed.